\documentclass[ars, final]{copernicus}
\nolinenumbers 

\begin{document}
\title{Using analog computers in today's largest computational challenges}

\Author[1]{Sven}{K\"oppel}
\Author[1]{Bernd}{Ulmann}
\Author[1]{Lars}{Heimann}
\Author[1,2]{Dirk}{Killat}

\affil[1]{Anabrid GmbH, Am Stadtpark 3, 12167 Berlin, Germany}
\affil[2]{Microelectronics Department, Brandenburg University of Technology, 03046 Cottbus, Germany}

\correspondence{Sven K\"oppel (koeppel@anabrid.com)}

\runningtitle{Using analog computers in today's largest computational challenges}
\runningauthor{K\"oppel, Ulmann, Heimann, Killat}

\received{}
\pubdiscuss{} 
\revised{}
\accepted{}
\published{}

\newcommand{\todo}[1]{\textcolor{blue}{TODO: #1}}

\newcommand{\ie}{\textit{i.e.},\ }
\newcommand{\cf}{\textit{c.f.},\ }


\firstpage{1}

\maketitle

\begin{abstract}
Analog computers can be revived as a feasible technology platform for low
precision, energy efficient and fast computing. We justify this statement by
measuring the performance of a modern analog computer and comparing it with
that of traditional digital processors. General statements are made about the solution
of ordinary and partial differential equations.
Computational fluid dynamics are discussed as an example of large scale scientific
computing applications. Several models are proposed which demonstrate the benefits
of analog and digital-analog hybrid computing.
\end{abstract}


\introduction  

Digital computing has transformed many --- if not close to all --- aspects
of industry, humanities and science. Turing completeness allows 
statements to be made about the computability and decidability of problems and \emph{computational
power} of machines. Digital storage has undergone numerous technological
advances and is available in increasingly vast amounts. Nevertheless, contemporary digital
computing is possibly not the last word in computing, despite its dominance in the consumer
market for the last 40+ years.

Fundamental research about non-traditional (also referred to as \emph{unconventional}
or \emph{exotic}) computing is taking place in material
sciences, chemistry but also in more exotic branches such as biology and life sciences.
Amongst others, beyond-Turing computing \citep{Siegelmann1995},
natural computing \citep{Calude99aglimpse},
neuromorphic computing  \citep{Schuman.5192017,Ziegler2020} or
quantum computing \citep{Zhou_2020,Georgescu_2014,Kenden2010} are fields of active
investigation. Being fundamental research at heart, these disciplines come with technological challenges.
For instance, computing with DNA still requires the use of large scale laboratory
equipment and machinery
\citep{DeatonGarzonRoseFranceschettiStevens1998}.
Currently, not only the low-temperature laboratory conditions but also the
necessary error correction schemes challenge practical quantum computers
\citep{BSIquantenComputer}.
This currently negates any practical advantage over silicon based digital computing.
Furthermore,
all of these alternative (or \emph{exotic}) computer architectures share
the characteristic that they are fundamentally \emph{non-portable}. This means they will
have to be located at large facilities and dedicated special-purpose computing centers for a
long time, if not forever. This is not necessarily a practical drawback,
since the internet allows for delocalization of systems.

In contrast to this, silicon based \emph{electronic analog computing} is a technology with
a rich history, 
which operates in a normal workplace environment
\citep[non-laboratory conditions; ][]{ap2}.
Digital computers overtook their analog counterparts in the last century,
primarily due to their ever-increasing digital clock speeds and their flexibility that
comes from their algorithmic approach and the possibility of using these machines in 
a time-shared environment. However, today Moore's law is coming to a hard
stop and processor clock speeds have not significantly increased in the past decade. Manycore
architectures and vectorization come with their own share of problems, given their
fundamental limits as described, for instance, by Amdahl's law \citep{AmdahlsLaw1985}.
GPGPUs and specialized digital computing chips concentrate on
vectorized, and even data flow-oriented programming paradigms but are still limited by
parasitic capacitances which determine the maximum possible clock frequency and provide a 
noticeable \emph{energy barrier}.

Thanks to their properties, analog computers have attracted the interest of many research groups.
For surveys of theory and applications, see for instance \citet{Bournez2018} or the
works of \citet{MacLennan2004,MacLennan2012,MacLennan2019}.
In this paper, we study the usability of analog computers for applications in science.
The fundamental properties of analog computers are low power requirements, low resolution
computation and intrinsic parallelism. Two very different uses cases/scenarios can be identified:
High performance computing (HPC) and low energy portable computing. The energy and
computational demands for both scenarios are diametrically-opposed and this paper is primarily
focused on HPC.

The paper is structured as follows: In Section~\ref{sec:linear}, we review the general
assumptions about digital and analog computing. In Section~\ref{sec:ode}, small scale benchmark
results are presented for a simple ordinary differential equation.
In Section~\ref{sec:pde}, a typical partial differential equation is considered as an 
example for a large
scale problem. Spatial discretization effects and computer architecture design choices are
discussed. Finally, Section~\ref{sec:summary} summarizes the findings.

\section{A Simple (Linear) Model for Comparing Analog and Digital Performance}\label{sec:linear}

In this paper, we study different techniques for solving differential equations computationally.
Due to the different conventions in algorithmic and analog approaches, a common language
had to be found and is described in this section. Here, the term \emph{algorithmic
approach} addresses the classical Euler method or classical quasi-linear techniques
in ordinary or partial differential equations (ODEs/PDEs), \ie  general methods of numerical mathematics.
The term \emph{analog approach} addresses the continuous time integration with
an operational amplifier having a capacitor in the feedback loop.
%
The fundamental measures of computer performance under consideration are the
{time-to-solution} $T$, the power consumption $P$ and the energy demand $E$.
\subsection{Time to solution}
The {time-to-solution} $T$ is the elapsed real time (lab time or wall clock time) for solving a
differential equation $\partial_t u = f(u)$ from its initial condition $u(t_0)$ at time
$t_0$ to some target simulation time $t_\text{final}$,
\ie for obtaining $u(t_\text{final})$. The \emph{speed factor} $k_0:=T/t_\text{final}$ is the ratio of
elapsed simulation time per wall clock time.
On analog computers, this allows to identify the \emph{maximum frequency} $\nu = k_0/(2\pi~\unit{sec})$.
On digital computers, the time-to-solution is used as an estimator (in a statistical sense) for the average $k_0$.
Relating this quantity to measures in numerical schemes is an important discussion point in this paper.
Given the simplest possible ODE,
\begin{equation}\label{eq:feqy}
  \mathrm{d\,}y/\mathrm{d\,}t=f(y):= y \,,
\end{equation}
one can study the analog/digital
computer performance in terms of the \emph{complexity} of $f(y)$. 
For a problem $M$ times as big as the given one, the inherently fully \emph{parallel} analog computer
exhibits a constant time-to-solution, \ie in other terms,
\begin{equation}
	T_A^M := T_A^1,
	\quad\text{\ie}\quad T_A^M = T_A^M(M) = \mathcal{O}(1),
\end{equation}
In contrast,
a single core (\ie nonvectorized, nor superscalar architecture) digital computer
operates in a \emph{serial} fashion and can achieve a time-to-solution 
\begin{equation}
	T_D^M = M\cdot T_D^1,
	\quad\text{\ie}\quad T_D^M = T_D^M(M)= \mathcal{O}(M).
\end{equation}
Here, $T^1$ refers to the time-to-solution for solving equation~\eqref{eq:feqy}, while 
$T^M$ refers to the time-to-solution for solving a problem $M$ times as hard.
$M \in \mathbb{N}$ is the measure for the algorithmic complexity of $f(y)$.
$f(M)=\mathcal{O}(g(M))$ refers to the Bachmann-Landau asymptotic notation.
%
%
The number of computational elements required to implement $f(y)$ on an analog computer or
the number of instructions required for computing $f(y)$ on a digital computer could provide
numbers for~$M$.
This is because it is assumed that the evaluation of $f(y)$ can hardly be numerically parallelized.
For a system of $N$ coupled ODEs $\mathrm{d\,} y_i/\mathrm{d\,}t=f_i(y_1,\dots,y_N)$, the vector-valued
$\vec f$ can be assigned an effective complexity $\mathcal{O}(NM)$ with the same reasoning.
However, an overall complexity $\mathcal{O}(M)$ is more realistic since parallelism could be
exploited more easily in the direction of $N$ (MIMD, multiple instruction, multiple data).

Furthermore, multi-step schemes
implementing higher order numerical time integration can exploit digital parallelization
(however, in general the serial time-to-solution of a
numerical Euler scheme is the limit for the fastest possible digital time integration).
Digital parallelization is always limited by the inherently serial parts of a problem 
\citep[{Amdahl's law}, ][]{AmdahlsLaw1985},
which makes the evaluation of $f(y)$ the hardest part of the problem. Section \ref{sec:pde}
discusses complex functions $f(y)$ in the context of the \emph{method of lines} for PDEs.

It should be emphasized that, in the general case, this estimate for the digital computer is
a most optimistic (best) estimate, using today's numerical methods.
It does not take into account hypothetical algorithmic
``shortcuts'' which could archive  solutions faster than $\mathcal{O}(M)$, because they imply some
knowledge about the internal structure of $f(y)$ which could probably also be exploited in analog
implementations.

\subsection{Power and energy scaling for the linear model}
For a given problem with time-to-solution $T$ and average power consumption $P$, the overall energy 
is estimated by $E=PT$ regardless of the computer architecture.

In general, an analog computer has to grow with the problem size $M$. Given constant power requirements per
computing element and neglecting increasing resistances or parasitic capacitances, in general one can
assume the analog computer power requirement $P_A^M$ for a size $M$ problem to scale from a size 1
problem $P_A^1$ as $P_A^M=P_A^1\cdot M$. In contrast, 
a serial single node digital computer 
in principle can compute a problem of any size serially by relying on dynamic memory (DRAM),
\ie $P_D^M = P_D^1$. That is, the digital computer 
power requirements for running a large problem ($P_D^M$) are (at first approximation) 
similar to running a small problem $P_D^1$. Typically,
the DRAM energy demands are one to two orders of magnitude smaller than those of a desktop or
server grade processor and are therefore negligible for this estimate. 

Interestingly, this model suggests that the overall energy requirements to solve a large
 problem on an analog and digital computer, respectively, are  both $E_D^M$ and $E_A^M = \mathcal{O}(M)$,
\ie the analog-digital energy ratio remains constant despite the fact that the analog
computer computes (runs) linearly faster with increasing problem size $M$. This can be easily
deduced by $E=P\cdot T$. In this model, it is furthermore
\begin{equation}\label{eq:eaed}
	\frac{E_A^M}{E_D^M} = \frac{M\, P_A^1}{P_D^1} \frac{T_A^1}{M\, T_D^1} = \frac{P_A^1}{P_D^1} \frac{T_A^1}{T_D^1} = \textrm{const}
	\,.
\end{equation}
The \emph{orthogonal} performance features of the fully-parallel analog computer
and the fully-serial digital computer are also summarized in Table \ref{table:linear-model}.

When comparing digital and analog computer power consumption, the power consumption under consideration should
include the \emph{total} computer power including administrative parts (like network infrastructure,
analog-to-digital converters or cooling) and power supplies. In this work, data of heterogenous sources are compared
and definitions may vary. 

\subsection{Criticism and outlook}
Given that the digital and analog technology (electric representation of information, transistor-based
computation) is quite similar, the model prediction of a similarly growing energy demand is useful.
Differences are of course hidden in the constants (prefactors) of the asymptotic notation $\mathcal{O}(M)$.
Quantitative studies in the next sections examine this prefactor in~$\mathcal{O}(M)$.

The linear model is already limited in the case of serial digital processors when the computation gets
\emph{memory bound} (instead of \emph{CPU-bound}). Having to wait for data leads to a performance drop and might
result in a worsened superlinear~$T_D^M$.

\emph{Parallel digital} computing as well as \emph{serial analog} computing
has not yet been subject of the previous discussion.
While the first one is a widespread standard technique, the second
one refers to analog-digital hybrid computing which, inter alia, allows a small analog
computer to be used repeatedly on a large problem, effectively rendering the analog part as an analog
accelerator or co-processor for the digital part. Parallel digital computing suffers from a theoretical
speedup limited due to the non-parallel parts of the algorithm 
\citep[see also][]{Gustafson1988},
which has exponential impact on $T_D^M$. This is where the intrinsically parallel analog computer exhibits
its biggest advantages. Section \ref{sec:pde} discusses this aspect of analog computing.

\begin{table}[t]
	\caption{A linear model for work: The computational cost $C$ of evaluating $f(u)$ in the ODE $\partial u/\partial t = f(u)$
		is expected to grow as $C \in \mathcal{O}(M)$. The effects on time-to-solution $T$, power $P$
		and energy $E$ demands are shown.
	}\label{table:linear-model}
	\begin{tabular}{lcc}
		\tophline
		(Quantity) & Digital & Analog   \\
		\middlehline
		$T(M)$ [s]                    & $\mathcal{O}(M)$	 & $\mathcal{O}(1)$  \\
		$P(M)$ [W]                    & $\mathcal{O}(1)$     & $\mathcal{O}(M)$  \\
		$E(M)$ [J]                    & $\mathcal{O}(M)$     & $\mathcal{O}(M)$   \\
		\bottomhline
	\end{tabular}
	\belowtable{} 
\end{table}

\section{A performance survey on solving ordinary differential equations (ODEs)}\label{sec:ode}

In this section, quantitative measurements between contemporary analog
and digital computers will be made. We use the \emph{Analog Paradigm Model-1}
computer \citep{Model1Handbook,ap2}, a modern modular academic analog computer and an ordinary
Intel\textsuperscript{\textcopyright} Whiskey Lake ``ultra-low power mobile'' processor (Core i7-8565U) as a
representative of a typical desktop-grade processor.
Within this experiment, we solve a simple\footnote{
 This equation is inspired by the \citet{dahlquist1979generalized} \emph{test equation} $y'=\lambda y$
 used for stability studies. The advantage of using an oscillator is the self-similarity of the
 solution which can be observed over a long time.
} test equation
$\mathrm{d\,}^2 y / \mathrm{d\,t}^2 = \lambda y$ (with real-valued $y$ and
$\lambda=\pm1$) on both a digital and analog computer.

\subsection{Time to solution}\label{sec:ODE-TTS}
The digital computer solved the simple ordinary differential equation (ODE) 
with simple text-book level scalar benchmark codes written in C and Fortran
and compiled with GCC.
Explicit (forward) integrator methods are adopted (Euler/Runge-Kutta).
The algorithm computed
$N=2\times 10^3$ timesteps with timestep size $\Delta t = 5\times 10^{-4}$ each
(see also section \ref{sec:pde} for a motivation for this time step size).
Therefore, it is $t_\text{final} = N\Delta t =1$. No output\footnote{Both in terms of \emph{dense output} or
  any kind of evolution tracking. A textbook-level approach with minimal memory footprint is adopted
  which could be considered an \emph{in-place} algorithm.
} was written during the benchmark to ensure
the best performance.
The time per element update (per integration step) was roughly $(45 \pm 35)\,\unit{ns}$.
For statistical reasons, the computation was repeated and averaged $10^5$ times.
Depending on the order of the integration scheme, the overall
wall clock time was determined as $T_D = (75\pm45)\,\unit{\mu s}$ in order to
achieve the simulation time $t_\text{final}$.

In contrast, the equation was implemented with integrating (and negating, if $\lambda=-1$) operational
amplifiers on the \emph{Analog Paradigm Model-1}. The machine approached 
$t_\text{final}=1$ in a wall-clock time $T_A = 1\,\unit{sec} / k_0$  with
$k_0 \in \{ 1, 10, 10^2, 10^3, 10^4 \}$ the available integration speed factors on the machine
\citep{Model1Handbook}.
The \emph{Analog Paradigm Model-1} reached the solution of $y''=y$ at $t_\text{final}=1$
in a wall-clock time $T_A=100\,\unit{\mu s}$ at best.

Note how $T_A/T_D \approx 1$, \ie in the case of the
smallest possible reasonable ODE, the digital computer (2020s energy efficient desktop
processor) is roughly as fast as the \emph{Analog Paradigm Model-1}
(modern analog computer with an integration level comparable to the 1970s).

Looking forward, given the limited increase in clock frequency, with a faster processor one can
probably expect an improvement of $T_D$ down to the order of $1\,\unit{\mu s}$.
For an analog computer on a chip, one can
expect an improvement of $T_A$ down to the order of $1\,\unit{\mu s}$-$10\,\unit{ns}$. 
This renders $T_A/T_D \approx 10^{-(1 \pm 1)}$ as a universal constant.

Summing up, with the given numbers above, as soon as the problem complexity grows, 
the analog computer outperforms the digital one, and this advantage increases linearly.

\begin{table}
	\caption{Small scaling summary: Measured benchmark (Intel\textsuperscript{\textcopyright}
		processor vs.  \emph{Analog Paradigm Model-1}) and expected/projected analog chip results.
	}\label{table:ode}
	\begin{tabular}{llll}
		\tophline
		           & \multicolumn{2}{c}{Measured}    & Projected \\
                     \cline{2-3}
		           & Digital & Analog (\emph{M1})    & Analog Chip  \\
		\middlehline
		$T$  [$\unit{\mu s}$]         & $75 \pm 45$          & $100$            & $10^{-(0.5 \pm 0.5)}$   \\ 
		$k_0 \sim 1/\Delta t$ [Hz]    & $3\times 10^4$          & $10^4$      & $10^{6.5 \pm 0.5}$  \\
		$P$ [$\unit{W}$]              & $10$                 & $0.4$            & $10^{-2}$  \\
		$E = P\cdot T$ [$\mu\unit{J}$] & $900 \pm 600$       & $40$             & $10^{-(2.5 \pm 0.5)}$  \\
		$F$ [FLOP/sec]                & $10^9$               & $3\times 10^{(4 \pm 1)}$  & $7 \times 10^5$ \\
		$F/E$ [FLOP/J]                & $10^8$               & $7.5\times 10^{8\pm 1}$   & $3\times 10^{11}$  \\
		%
		\bottomhline
	\end{tabular}
	\belowtable{} 
\end{table}

\subsection{Energy and power consumption}
The performance measure codes
\texttt{likwid}~ \citep{Likwid,Likwid2,LikwidDoi} 
and \texttt{perf} \citep{perf1}
were used in order to measure 
the overall floating-point operations (FLOP) and energy usage of the digital processor.
For the Intel mobile processor, this provided a power consumption of $P_D=10\,\unit{W}$ during computing.
This number was derived directly from the CPU performance counters. 
The overall energy requirement was then $E_D = P_D T_D = (0.9 \pm 0.6)\,\unit{mJ}$.
Note that this number only takes the processor energy demands into account, not any other auxiliary
parts of the overall digital computer (such as memory, main board or power supply). For the overall
power consumption, an increase of at least 50\% is expected.

The analog computer energy consumption is estimated as $P_A\approx 400\,\unit{mW}$.
The number is based on measurements of actual \emph{Analog Paradigm Model-1} computing units,
in particular $84\,\unit{mW}$ for a single
summer and $162\,\unit{mW}$ for a single integrator. The overall energy requirement is then
$E_A = P_A T_A = 40\,\unit{\mu J}$.

Note that $P_D/P_A \approx 25$, while $E_D/E_A \approx (2.25 \pm 1.5)$. The conclusion is that the
analog and digital computer require a similar amount of energy for the given computation,
a remarkable result given the 40-year technology gap between the two architectures compared here.


For power consumption, it is hard to give a useful projection due to the accumulating administrative
overhead in case of parallel digital computing, such as data transfers, non-uniform memory
accesses (NUMA) and switching networking infrastructure. It can be assumed that this will change
the ratio $E_D/E_A$ further in favor of the analog computer for both larger digital and 
analog computers. Furthermore, higher integration levels lower $E_A$: the
\emph{Analog Paradigm Model-1} analog computer is realized with an integration level comparable
with 1970s digital computers. We can reasonably expect a drop of two to three orders of
magnitude in power requirements with fully integrated analog computers.

\subsection{Measuring computational power: FLOP per Joule}
For the digital computer, the number of computed \emph{floating-point operations} (FLOP\footnote{
   sic! We either argue with \emph{overall} FLOP and Energy (Joule) or \emph{per second} quantities such as
   FLOP/sec (in short FLOPS) and Power (Watt). In order to avoid confusion, we avoid the abbreviation 
   ``FLOPS'' in the main text.
   Furthermore, SI prefixes are used, \ie $\unit{kFLOP} = 10^3\,\unit{FLOP}$,
   $\unit{MFLOP} = 10^6\,\unit{FLOP}$ and $\unit{GFLOP} = 10^9\,\unit{FLOP}$.
}) can be measured.
The overall single core nonvectorized performance was measured as $F \approx 1\,\unit{GFLOP/sec}$.
A single computation until $t_\text{final}$ required roughly $F_D=3\,\unit{kFLOP}$.
The ratio $F_D/P_D = 100\,\unit{MFLOP/J}$ is a measure of the number of computations per
energy unit on this machine. This performance was one to two orders less than typical HPC numbers.
This is because an energy-saving desktop CPU and not a high-end processor was benchmarked.
Furthermore, this benchmark was by purpose single-threaded.

In this non-vectorized benchmark, the reduced resolution of the analog computer was ignored.
In fact it is slightly lower than an IEEE 754 half precision
floating-point, compared to the double precision floating-point numbers in the digital benchmark.
One can then assign the analog computer a \emph{time-equivalent} floating-point operation performance
\begin{equation}
   F_A := F_D \frac{T_A}{T_D} \approx 10^{(1 \pm 1)}F_D = 3\times 10^{(4 \pm 1)}\,\unit{FLOP} \,.
\end{equation}
The analog computer FLOP-per-Joule ratio (note that $\unit{FLOP/J} = \unit{FLOPs/W}$) is
\begin{equation}
\frac{F_A}{E_A} = \frac{3\times 10^{(4 \pm 1)} \unit{FLOP}} {40 \,\unit{\mu J}} = 7.5 \times 10^{8 \pm 1} \,\unit{FLOP/J} \,.
\end{equation}
That is, the analog computer's ``FLOP per Joule'' is slightly larger than for the digital one.
Furthermore, one can expect an increase of $F_A/E_A$ by 10-100 for an analog computer chip.
See for instance
\citet{CowanDiss2005} and \citet{Cowan2005AVA}, who claim $20\,\unit{GFlop/sec}$. We expect $300\,\unit{GFlop/sec}$
to be more realistic, thought (Table \ref{table:ode}).

Keep in mind that the FLOP/sec or FLOP/J measures are (even in the case of comparing two digital computers)
always problem/algorithm-specific (\ie in this case a Runge Kutta solver of  $y''=y$) and therefore
controversial as a comparative figure.
\section{PDEs and many degrees of freedom}\label{sec:pde}
This section presents forecasts about the solution of large scale differential equations. No benchmarks
have been carried out, because a suitable integrated analog computer on chip does not yet exist.
For the estimates, an analog computer on chip with
an average energy consumption of about $P_N=4\,\unit{mW}$ per computing element
(\ie per integration, multiplication, etc.) 
and maximum frequency $\nu = 100\,\unit{Mhz}$, which is refered to as the \emph{analog} maximum frequency 
$\nu^A$ in the following, was assumed.was assumed.\footnote{Summation will be done implicitly on chip by
	making use of Kirchhoff's law (current summing) so that no explizit computing element are required for this operation.}
These numbers are several orders of magnitude
better than the $P_N=160\,\unit{mW}$ and $\nu = 100\,\unit{kHz}$ of the \emph{Analog Paradigm Model-1} computer discussed
in the previous section. For the digital part, different systems than before are considered.

In general,
the bandwidth of an analog computer depends on the frequency response characteristics of the elements,
such as summers and integrators. The actual achievable performance also depends on the technology. 
A number of examples shall be given to motivate our numbers:
In $65\,\unit{nm}$ CMOS technology, bandwidths of over $2\,\unit{GHz}$ are achievable with integrators 
\citep{Breems2016}.
At unity-gain frequencies of $800\,\unit{MHz}$ to $1.2\,\unit{Ghz}$ and power consumption
of less than $2\,\unit{mW}$, integrators with a unity-gain frequency of $400\,\unit{Mhz}$ are achievable
\citep{Wang2018}.
%

\subsection{Solving PDEs on digital and analog computers}
Partial differential equations (PDEs) are among the most important and powerful mathematical frameworks
for describing dynamical systems in science and engineering.
PDE solutions are usually fields $\vec u = \vec u(\vec r, t)$,
\ie functions\footnote{The explicit dependency on $\vec r$ and $t$ is omitted in the following
	text.} of spatial position $\vec r$ and time~$t$. In the following, we concentrate on initial
value boundary problems (IVBP). These problems are described by a set of PDEs valid within a 
spatial and temporal domain and complemented with field values imposed on the domain boundary.
For a review of PDEs, their applications and solutions see for instance
\citet{Brezis1998}.
In this text, we use computational fluid dynamics (CFD) as a representative theory for
discussing general PDE performance. In particular, classical
hydrodynamics (Euler equation) in a flux-conservative formulation
is described by hyperbolic conservation laws in the next sections.
Such PDEs have a long tradition of being solved with highly accurate numerical schemes.

Many methods exist for the spatial discretization. While finite volume schemes are popular for their
conservative properties, finite difference schemes are in general cheaper to implement. In this
work, we stick to simple finite differences on a uniform grid 
with some uniform grid spacing $\Delta \vec r$.
The evolution vector field $\vec u(\vec r,t)$ is sampled on $G$ grid points per dimension and
thus replaced by $\vec u_k(t)$ with $0 \leq k < G$. It is worthwhile to mention that this approach
works in classical orthogonal ``dimension by dimension'' fashion, and the number of total grid points
is given by $G^{D}$. The computational domain is thus bound by
$\Omega = [\vec r_0, \vec r_{G}]^D$.
A spatial derivative $\partial_i f$ is then approximated by a central finite difference scheme,
for instance $\partial_i f_k \approx (f_{k+1} - f_{k-1})/(2\Delta x) + \mathcal O(\Delta x^2)$
for a second order accurate central finite difference approximation of the derivative of
some function $f$ at grid point~$k$.

Many algorithmic solvers implement numerical schemes which exploit the
\emph{vertical method of lines} (MoL) to rewrite the PDE into coupled ordinary differential equations
(ODEs). Once applied, the ODE system can be written as $\partial_t u^k = G^k(\vec u, 
\vec \nabla \vec u)$ 
with $u^k$ denoting the time evolved (spatial) degrees of freedom and $G^k$ functions 
containing spatial derivatives ($\partial_i u^j$) and algebraic sources. A standard
time stepping method determines a solution $u(t_1)$ at later time $t_1 > t_0$ by basically
integrating $u^k(t_1) = \int_{t_0}^{t_1} G^k(\vec u(t)) \mathrm d\, t + u^k(t_0)$.
Depending on the details of the scheme, $G^k$ is evaluated (probably repeatedly or in a
weak-form integral approach) during the time integration of the system. However, note
that other integration techniques exist, such as the arbitrary high order 
ADER technique \citep{Titarev2002,TITAREV2005715}.
The particular spatial discretization method has a big impact on the computational cost
of $G^i$. Here, we focus on the (simplest) finite difference technique,
where the number of neighbor communications per dimension grows linearly with the
\emph{convergence order} of the scheme.

\subsection{Classical Hydrodynamics on analog computers} \label{sec:hydro-definitions}

The broad class of \emph{fluid dynamics} will be discussed
as  popular yet simple type of PDEs.
It is well known for its efficient description
of the flow of liquids and gases in motion and is applicable in many domains such as
aerodynamics, in life sciences as well as fundamental sciences
\citep{Sod1985,Yih79,Yi19}. In this text, the simplest formulation is investigated: the
Newtonian hydrodynamics (also refered to as \emph{Euler equations}) with an ideal gas
equation of state. It is given by a nonlinear
PDE describing the time evolution of a mass density $\rho$, it's
velocity $v^i$, momentum $p^i=\rho v^i$ and energy $e=t+\varepsilon$, with the
kinetic contribution $t = \rho~\vec v^2 / 2$ and an ``internal'' energy $\varepsilon$, which
can account for forces on smaller length scales than the averaged scale.

Flux conservative \emph{Newtonian hydrodynamics} with an ideal gas equation of state
are one of the most elementary and text-book level formulations of fluid dynamics
\citep{Toro1998,harten1997high,hirsch1990numerical}. The
PDE system can be written in a dimension agnostic way in $D$ spatial dimensions
(\ie independent of the particular choice for $D$) as
\begin{equation} \label{eq:hydro}
	\frac{\partial \vec u}{\partial t} - \vec \nabla \cdot \vec f = \vec S
	~~\text{with }~\vec \nabla \cdot \vec f = \sum_i^{n_d} 	\frac{\partial \vec f^i}{\partial x^i} \,,
\end{equation}
\begin{equation}
	\vec u = \begin{pmatrix} \rho \\ p^j \\ e \end{pmatrix}
	~~\text{, }~
	\vec f^i = \vec f^i(\vec u, \vec v) = \begin{pmatrix} p^i \\  p^i v^j - p\, \delta^{ij} \\ v^i~(e+p) \end{pmatrix},
\end{equation}
with $i,j \in [1..D]\,.$
Here, the pressure $p=\rho\, \varepsilon(\Gamma-1)$ defines the ideal gas equation of state,
with adiabatic index $\Gamma=2$ and $\delta^{ij}$ is the Kronecker delta.
A number of vectors are important in the following: The integrated state
or \emph{evolved} vector $\vec u$ in contrast to the primitive state vector or
\emph{auxiliary} quantities $\vec v(u) = (p, v^i)$, which is a collection of
so called \emph{locally reconstructed} quantities.
Furthermore,
the right hand sides in \eqref{eq:hydro} do not explicitly depend on the spatial derivative
$\partial^i \rho$, thus the conserved flux vector $\vec f = \vec f(\nabla\vec q, \vec v)$ is
only a function of the derivatives of the \emph{communicated} quantities
$\vec q = (e, p^i)$ and the auxiliaries $\vec v$.
Furthermore, $\vec q$ and $\vec v$ are both functions of $\vec u$ only.

$\vec S=0$ is a source term. Some hydrodynamical models can be coupled by purely choosing
some nonzero $\vec S$, such as the popular \emph{Navier Stokes} equations which describe
viscous fluids. Compressible Navier Stokes equations can be written with a source term
 $\vec S = \vec \nabla \cdot \vec F^v$, with
\begin{align}
&\text{diffusion fluxes}~~ \textstyle \vec F^v = (0, \tau^{ij}, \sum_k \tau^{ik} v^k - q^j)^T, \\
&\text{viscous stress}~~ \textstyle \tau^{ij} = \mu ( \partial^i v^j + \partial^j v^i - \frac 23 (\partial^k v^k) \delta^{ij} ), \\
&\text{and heat flux}~~ \textstyle q^i = - (c_p \mu / Pr) \partial^i T,
\end{align}
with specific heats $c_p$, $c_v$, viscosity coefficient $\mu$,
Prandtl number $Pr$ and temperature $T$ determined by the perfect gas equation of state, \ie 
$T=(e-\vec v^2)/(2c_v)$.
The computational cost from Euler equation to Navier Stokes equation is roughly \emph{doubled}.
Furthermore, the partial derivatives on the velocities and temperatures also \emph{double}
the quantities which must be communicated with each neighbor in every dimension. 
We use Euler equations in the following section for the sake of simplicity.

\begin{figure*}
	\includegraphics[width=\linewidth]{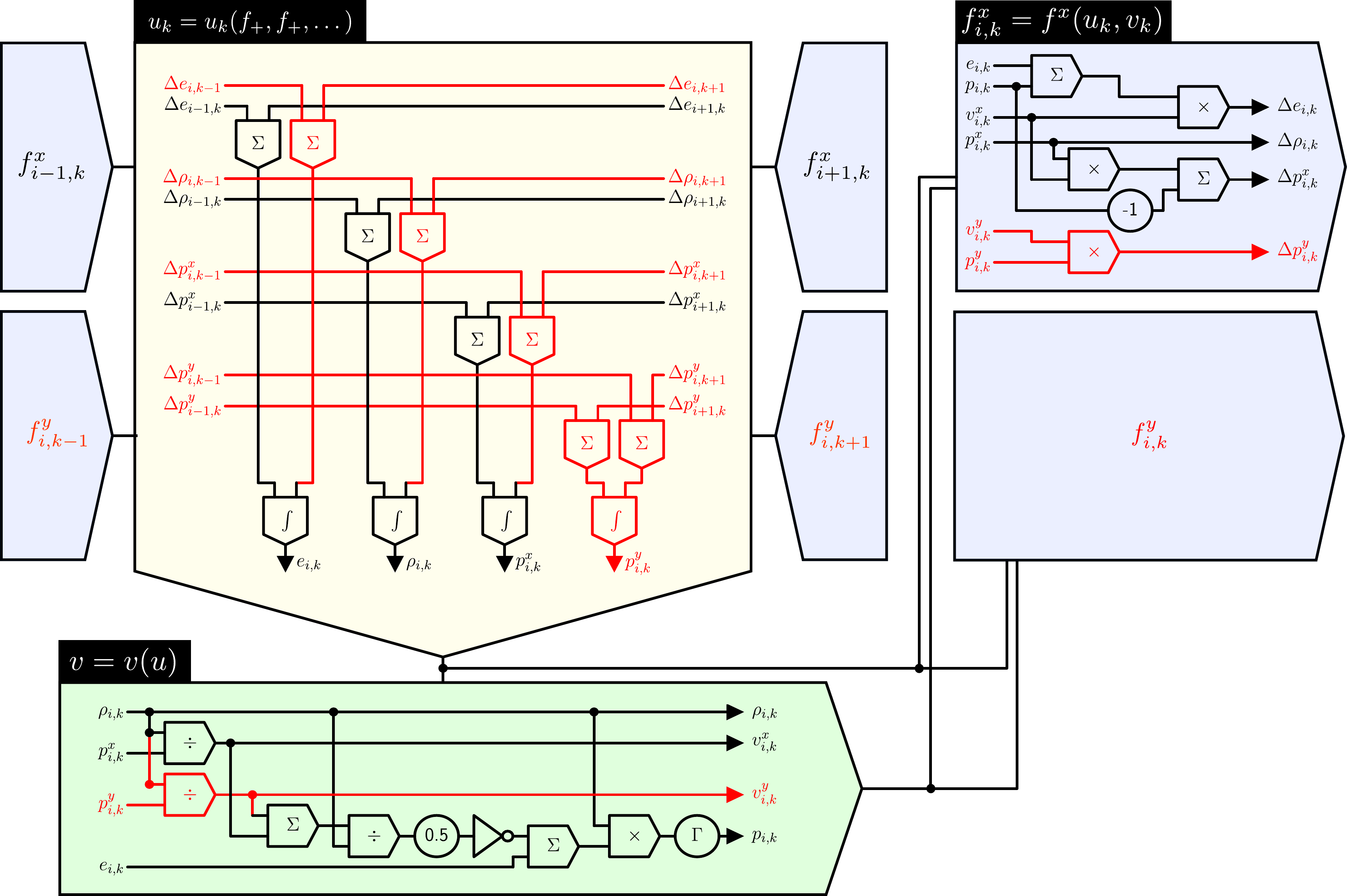}
	\caption{Overview  circuit showing the blocks $\vec f$, $\vec u$ and $\vec v$.
		The three labeled blocks are distinguished by colour. Information flow is
		indicated with arrows. The overall circuit is given for lowest order (RK1)
		and in one spatial dimension. The red circuitry is the required addition for
		two spatial dimensions. All computing elements are drawn ``abstractly'' and
		could be directly implemented with (negating) operational amplifiers on a
		very large \emph{Analog Paradigm Model-1} analog computer.
	}
	\label{fig:hydro-circuit}
\end{figure*}
\subsection{Spatial discretization: Trading interconnections vs. computing elements}
Schemes of (convergence) order $F$ shall be investigated, which require the communication
with $F$ neighbour elements.
For instance, a $F=4$th order accurate stencil has to communicate and/or
compute four neighbouring elements $\vec f_{k-2}, \vec f_{k-1}, \vec f_{k+1}, \vec f_{k+2}$.
Typically, long-term evolutions are carried out with $F=4$ or $F=6$.
In the following, for simplicity, second order stencil ($F=2$) is chosen.
One identifies three different subcircuits 
\begin{equation}
\vec u_k(\vec f_{k+1}, \vec f_{k-1}) := \int \left( \vec f_{k+1} - \vec f_{k-1}  \right)~ \mathrm{d\,t} / (2\Delta x) \,,
\end{equation}
with
$ \vec f_{k\pm 1} := \vec f_k(\vec q_{k \pm 1}, \vec v_k)$
and 
$ \vec v_k := \vec v_k(\vec u_k) $
according to their previous respective definitions.
Figure \ref{fig:hydro-circuit} shows this ``building block'' for a {single} grid point, an exemplar
for up to $D=2$ dimensions with an $F=2$nd order finite difference stencil. The circuit identifies a
number of intermediate expressions which are labeled as these equations:
{ \small 
\begin{align}\label{eq:hydro-written-out}
&
{\partial_t \begin{pmatrix} \rho_{i,k} \\ p^x_{i,k} \\ \textcolor{red}{p^y_{i,k}} \\ e_{i,k} \end{pmatrix}}
\\ \nonumber
&= 
\frac{
\overbrace{\begin{pmatrix} p^x_{i+1,k} \\ p^x_{i+1,k} v^x_{i+1,k} - p_{i+1,k} \\ \textcolor{red}{p^x_{i+1,k} v^y_{i+1,k}} \\ v^x_{i+1,k} (e_{i+1,k}+p_{i+1,k}) \end{pmatrix}}
  ^{\vec f^x_{i+1,k}}
- 
\overbrace{\begin{pmatrix} p^x_{i-1,k} \\ p^x_{i-1,k} v^x_{i-1,k} - p_{i-1,k} \\ \textcolor{red}{p^x_{i-1,k} v^y_{i-1,k}} \\ v^x_{i-1,k} (e_{i-1,k}+p_{i-1,k}) \end{pmatrix}}
  ^{\vec f^x_{i-1,k}}
}{2\Delta x}
\\ \nonumber
&+
\textcolor{red}{
\overbrace{
\frac{\begin{pmatrix} p^y_{i,k+1} \\ p^y_{i,k+1} v^x_{i,k+1} \\ p^y_{i,k+1} v^y_{i,k+1} - p_{i,k+1} \\ v^y_{i,k+1} (e_{i,k+1}+p_{i,k+1}) \end{pmatrix}}{2\Delta y}
}^{\vec f^y_{i,k+1}}
-
\overbrace{
\frac{\begin{pmatrix} p^y_{i,k-1} \\ p^y_{i,k-1} v^x_{i,k-1} \\ p^y_{i,k-1} v^y_{i,k-1} - p_{i,k-1} \\ v^y_{i,k-1} (e_{i,k-1}+p_{i,k-1}) \end{pmatrix}}{2\Delta y}
}^{\vec f^y_{i,k-1}}
}
\\ \nonumber
&=
\frac{
\begin{pmatrix} \Delta \rho_{i+1,k} \\ \Delta p^x_{i+1,k}	 \\ \textcolor{red}{\Delta p^y_{i+1,k}} \\ \Delta e_{i+1,k} \end{pmatrix}
-
\begin{pmatrix} \Delta \rho_{i-1,k} \\ \Delta p^x_{i-1,k}	 \\ \textcolor{red}{\Delta p^y_{i-1,k}} \\ \Delta e_{i-1,k} \end{pmatrix}
}{2\Delta x}
\textcolor{red}{
+
\frac{
	\begin{pmatrix} \Delta \rho_{i,k+1} \\ \Delta p^x_{i,k+1}	 \\ \Delta p^y_{i,k+1} \\ \Delta e_{i,k+1} \end{pmatrix}
	-
	\begin{pmatrix} \Delta \rho_{i,k-1} \\ \Delta p^x_{i,k-1}	 \\ \Delta p^y_{i,k-1} \\ \Delta e_{i,k-1} \end{pmatrix}
}{2\Delta y}
}
\end{align}
}
Just like in Figure~\ref{fig:hydro-circuit}, all expressions which are vanishing in a single spatial dimension are
colored in red. Furthermore, note how the index $i$ denotes the $x$-direction and $k$ the $y$-direction,
and that there are different fluxes $\vec f^j$ in the particular directions. \eqref{eq:hydro-written-out}
is closed with the element-local auxiliary recovery
\begin{equation}
\begin{pmatrix}
	v^x_{i,k} \\ \textcolor{red}{ v^y_{i,k}} \\ p_{i,k}
\end{pmatrix}
= \begin{pmatrix}
	p^x_{i,k} / \rho_{i,k} \\ \textcolor{red}{ p^y_{i,k} / \rho_{i,k}} \\
    e_{i,k} - \rho_{i,k} \left( {(v^x_{i,k})^2 + \textcolor{red}{(v^y_{i,k})^2}}  \right)/2
\end{pmatrix} \,.
\end{equation}

Note that one can trade neighbor communication (\ie number of wires between grid points)
for local recomputation. For instance, it would be mathematically clean to communicate only the
conservation quantities $\vec u$ and reconstruct $\vec v$ whenever needed. In order to
avoid too many recomputations, some numerical codes also communicate parts of $\vec v$. In
an analog circuit, it is even possible to communicate parts of the finite differences,
such as the $\Delta \vec v_{i,k}$ quantities in equation~\eqref{eq:hydro-written-out}.

The number of analog computing elements required to solve the Euler equation
on a single grid point is determined as $N_\text{single} = 5D + 5F(D+2) + 9$,
with $D$ being the number of spatial dimensions 
and $F$ the convergence order (\ie basically the finite
difference stencil size).
Typical choices of interest are convergence orders of $F\in[2,6]$ in $D\in[1,3]$ spatial dimensions.
Inserting the averaged $F=3\pm1$ and $D=2\pm 1$ into $N_\text{single}$ yields an averaged
$N_\text{single} \approx (84 \pm 40)$ computing elements per spatial degree of freedom
(grid point) required for implementing Euler equations.

Unfortunately, this circuit is too big to fit on the \emph{Analog Paradigm Model-1} computer resources
available. Consequently the following discussion is based on a future implementation using a large number of
interconnected analog chips. It is noteworthy that this level of integration is necessary to implement
large scale analog computing applications. With $P_N=4\,\unit{mW}$ per computing element,
the average power per spatial degree of freedom (\ie single grid point) is $P_{ND} = (336\pm160)\,\unit{mW}$.
\subsection{Time to solution}

\begin{figure*}[t!]
	\includegraphics[scale=0.63]{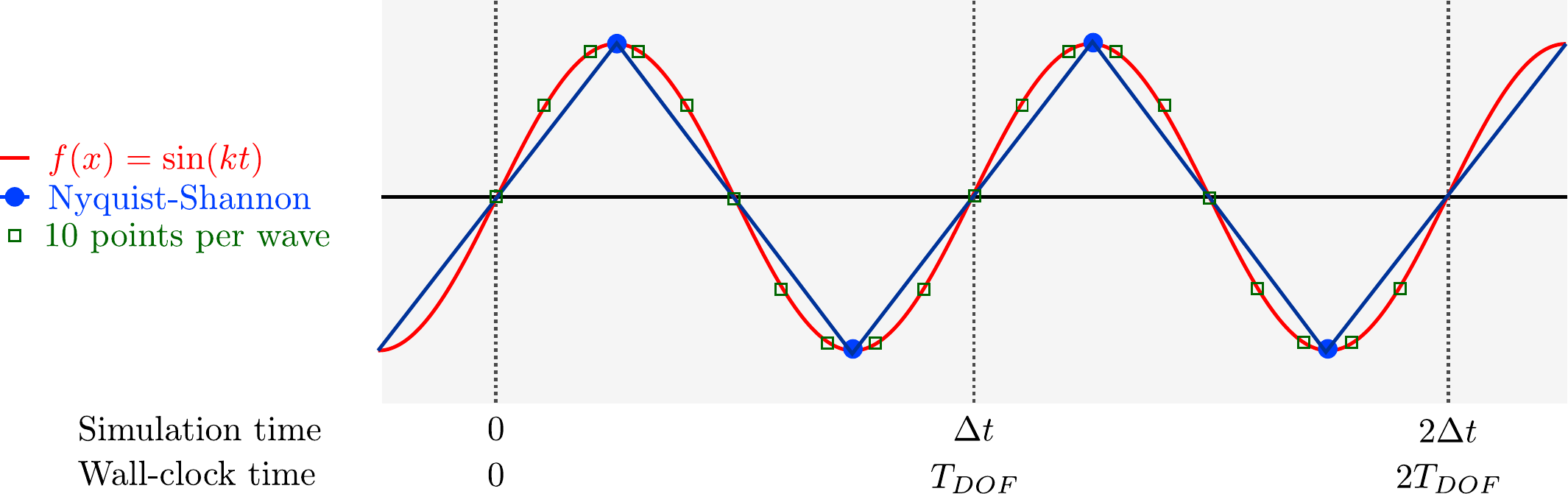}
	\caption{Analog signal sampling vs. numerical time integration: The time evolved
		sine with \emph{maximum frequency} $k=2\pi/\lambda$ has just the wavelength $\lambda=\Delta t$,
		with $\Delta t$ being the timestep size of the explicit Euler scheme. The Nyquist-Shannon theorem
		allows to determine wave length and phase position with two sampling points per wave length.
		However, a first order reconstruction of numerical data shows a triangle (zigzag) function. In contrast, the
		full wave is clearly visible at analog integration. More sampling points close the gap
		between analog and numerical representation.
	}
	\label{fig:cutoff-frequency}
\end{figure*}
Numerical PDE solvers are typically benchmarked using a \emph{wall-clock time per degree
of freedom update} measure $T_\text{DOF}$, where \emph{element update} typically means a time integration timestep.
In this measure, the overall wall clock time is normalized (divided) by the number of spatial
degrees of freedom as well as the number of parallel processors involved.

The fastest digital integrators found in literature carry out a time per degree of freedom
update $T_\text{DOF} = 10^{1\pm 1}\,\unit{\mu s}$. Values smaller
than $1\,\unit{\mu s}$ require already the use of sophisticated communication avoiding
numerical schemes such as discontinuous Galerkin (DG) schemes.\footnote{$h-p$ methods, which provide both
mesh refinement in grid spacing $h$ as well as a ``local'' high order description typically
in some function base expansion of order $p$. For reviews, see for instance
\citet{Cockburn2001} or \citet{SHU2016598}.} For instance,
\citet{Dumbser2008} demonstrate the superiority of so called $P_N P_M$
methods (polynomial of degree $N$ for reconstruction and $M$ for
time integration, where the limit $P_0 P_M$ denotes a standard high-order finite volume scheme)
by reporting $T_\text{DOF} = 0.8\,\unit{\mu s}$ for a $P_2 P_2$ method when solving two-dimensional Euler
equations. \citet{Diot2012} report an adaptive scheme which performs no faster than 
$T_\text{EU} = 30\,\unit{\mu s}$ when applied to three-dimensional Euler equations.
The predictor-corrector arbitrary-order ADER scheme applied by \citet{Koeppel2018} and \citet{Fambri2018}
to the \emph{general-relativistic magnetodynamic} extension of hydrodynamics reported $T_\text{DOF} = 41\,\unit{\mu s}$
as the fastest speed obtained. The non-parallelizable evaluation of more complex hydrodynamic models is clearly
reflected in the increasing times $T_\text{DOF}$.

Recalling the benchmark result of $T_\text{DOF} \sim 45\,\unit{ns}$ from Section \ref{sec:ODE-TTS},
the factor of 1000 is mainly caused by the inevitable communication required for obtaining
neighbor values when solving $f(y,\nabla y)$ in $\partial_t y = f(y)$. Switched networks have an
intrinsic communication latency and one cannot expect $T_\text{DOF}$ to shrink significantly, even for
newer generations of supercomputers. A key advantage of analog computing is that grid neighbor communication
happens continuously in the same time as in the grid-local circuit. That is, no time is lost for communication.

One can do a comparison with the analog computer without knowing the simulation time step size $\Delta t$.
The reasoning is based on the maximum frequency, \ie the shortest wavelength which can be resolved with
a (first order in time\footnote{
 For a high order time integration scheme, the cutoff increases formally linearly as
 $f_0\sim p/(10T_\text{DOF})$.
 That is, for a fourth order scheme, the digital computer is effectively four times faster in
 this comparison.
}) numerical scheme is $f_\text{sim} := 1/(10 \Delta t)$, \cf Figure \ref{fig:cutoff-frequency}.
The factor $10 = 2 \cdot 5$ includes a factor of
$2$ due to the Nyquist-Shannon sampling theorem, while the factor of $5$ is chosen to take into account that
a numerical scheme can marginally reconstruct a wave at frequency $f=1/(2\Delta t)$ by two points while it
can be obtained perfectly by the analog computer (down to machine precision without any artifacts). The
integration of signals beyond the maximum frequency results in a nonlinear response which heavily depends on the
electrical details of the circuit (which are beyond the scope of the analog computer architecture discussed in
this paper). One can demand that the numerical integrator time resolution is good enough to reconstruct a signal
\emph{without} prior knowledge on the wave form even at the maximum frequency.\footnote{
  Note that on a digital computer, the maximum frequency is identical to a \emph{cutoff frequency}
  (also refered to as \emph{ultraviolet cutoff}). On analog computers, there is no such \emph{hard}
  cutoff as computing elements tend to be able to compute with decreased quality at higher frequencies.
} This drives the demand for $5$ additional
sampling points per half-wave, in order to make analog and digital outcome comparable (see also figure \ref{fig:cutoff-frequency}).

It is noted that this argument is relevant as long as one is interested in obtaining and preserving the \emph{correct} time
evolution (of a system described by the differential equation) with an analog or digital computer, respectively.
In general, it is not valid to reduce the computational correctness within the solution domain of an initial value problem as
this will invalidate any later solution.

By assigning the numerical PDE solver a maximum frequency identical to the highest frequency which can be 
evolved by the scheme in a given time, one introduces an \emph{effective digital computer maximum frequency}
\begin{equation}
	\nu^D=1/(10 T_\text{DOF}) = 10^{1\pm 1}\,\unit{kHz} \,.
\end{equation}
Note how the mapping of simulation time (interval) $\Delta t$ to
wall-clock time (interval) $T_\text{DOF}$ results in a mapping of
simulation frequency $f_\text{sim}$ to wall-clock (or real-time) frequency
$\nu^D$ (Figure \ref{fig:cutoff-frequency}).

The calculated $\nu^D=10^{-2\pm 1}\,\unit{MHz}$ has to be contrasted with $\nu^A=100\,\unit{MHz}$ of the analog computer chip. One can conclude that
analog computers can solve large scale high performance computing at least $\nu^A/\nu^D=10^{3 \pm 1}$ times faster
than the digital ones, when $T_A$ and $T_D$ are the analog and digital time to solution.
Since $T\sim 1/\nu$, the resolution time reduces accordingly and $T_A/T_D = 10^{-3\pm1}$.

This is a remarkable result as it already assumes the fastest numerical integration schemes on a
\emph{perfectly} scaling parallel digital computer. In practical problems, these assumptions are hardly
ever met: The impossibility of (ideal) parallelization is one of the major drawbacks of digital computing. Nevertheless,
the above results show that even without these drawbacks, the analog computer is orders of magnitude faster.
Notably, while it needs careful adjustment both the problem and the code for a high-performance computer to achieve
acceptable parallel performance, when using an analog computer these advantages come effortless. The only
way to reduce the speed or timing advantage is to choose a disadvantegeous or unsuitable number scaling.

In this study the low resolution of an analog computer
(which is effectively IEEE 754 half precision floating-point) has been neglected.
In fact, high order time integration schemes
can invest computing time in order to achieve machine level accuracy which a typical error
$\Delta f_\text{digital}  \sim 10^{-10}$
on some evolved function or field $f$ and an error definition
$\Delta f_\text{simulation}:= (f_\text{simulation} - f_\text{exact}) / f_\text{exact}$.
An analog computer is limited by its intrinsic accuracy with
a typical error $\Delta f_\text{analog} \sim 10^{-(4\pm 1)}$
(averaging over the \emph{Analog Paradigm Model-1} and future analog computers on chip).

\subsection{Energy and power consumption}

One expects the enormous speedup $T_A/T_D$ of the analog computer to result in a much lower energy
budget $E_D = (T_D / T_A) E_A  \allowbreak = 10^{3 \pm 1} E_A$ for a given problem. However, as the power requirement
is proportional to the analog computer size, $P_A = N P_{ND}$, the problem size (number of grid points)
which can be handled by
the analog computer is limited by the overall power consumption. 
For instance, with a typical high performance computer power consumption of $P_A = 20\,\unit{MW}$,
one can simultaneously evolve a grid with $N=P_A/P_{ND}= 10^{11\pm0.5}$ points. This is in the same order of magnitude
as the largest scale computational fluid dynamics simulations evolved on digital high performance computer
clusters \citep[\cf Green 500 list,][]{green500_1,green500_2}. Note that in such a setup, the solution
is obtained on average $10^{3\pm1}$ times faster with a purely analog computer and consequently also the
energy demand is $10^{3\pm1}$ times lower.

Just to depict an analog computer of this size: Given 1000 computing elements per chip, 1000 chips per rack unit,
40 units per rack still requires 2,500 racks to build such a computer in a traditional design. This is one order of 
magnitude larger than the size of typical high performance centers. Clearly, at such a size the interconnections will
also have a considerable power consumption, even if the monumental engineering challenges for such a large scale
interconnections can be met. On a logical level, interconnections are mostly wires and switches (which require
little power, compared to computing elements). This can change dramatically with level converters and an energy
estimate is beyond the scope of this work.

\subsection{Hybrid techniques for trading power vs. time}

The analog computers envisaged so far have to grow with problem size (\ie with grid size, but also with equation complexity).
Modern chip technology could make it
theoretically possible to build a computer with $10^{12}$ analog computing elements, which is many orders of
magnitude larger than any analog computer that has been built so far (about $10^3$ computing elements
at maximum). The idea of combining
an analog and a digital computer thus forming a \emph{hybrid computer}
featuring analog and digital computing elements is not new. 
With the digital memory and algorithmically controlled program flow, a small analog computer
can be used repeatedly on a larger problem under control of the digital computer it is mated to.
Many attempts at solving
PDEs on hybrid computers utilized the analog computer for computing the
element-local updated state with the digital computer looping over the spatial
degrees of freedom. In such a scheme, the analog computer fulfils the role
of an \emph{accelerator} or \emph{co-processor}. Such attempts are subject of various
historical \citep[such as][]{nomura1968,reihing1959,vichnevetsky1968,vichnevetsky1971,volynskii1965,
	bishop1970,karplus1971,feilmeier1974}
and contemporary studies \citep[for instance][]{amant2014,Huang2017}.

A simple back-of-the-envelope estimation with a modern hybrid computer tackling the
$N=10^{11}$ problem is described below. The aim is to trade the sheer number of computing elements with their
electrical power $P$, respectively, against solution time $T$.
It is assumed that the analog-digital hybrid scheme works similarly to numerical parallelization:
The simulation domain with $N$ degrees of freedom is divided into $Q$ parts which can be evolved independently to a certain
degree (for instance in a predictor-corrector scheme). This allows to use a smaller analog
computer which only needs to evolve $N/Q$ degrees of freedom at a time. While the power consumption of such a
computer is reduced to $P_A \to P_A/Q$, the time to solution increases to $T_A \to Q T_A$. Of course, the
overall required energy remains the same, $E_A = P_A T_A = (P_A/Q) (Q T_A)$.

In this simple model, energy consumption of the digital part in the hybrid computer as well as numerical details
of the analog-digital hybrid computer scheme have been neglected. This includes the time-to-solution overhead introduced
by the numerical scheme implemented by the digital computer (negligible for reasonably small $Q$) and the power demands of 
the ADC/DAC (analog-to-digital/digital-to-analog) converters (an overhead which scales with $(D+2)G^D/Q$, \ie the state vector
size per grid element).

Given a fixed four orders of magnitude speed difference $\nu^D/\nu^A=10^4$ and a given physical problem
with grid size $N=10^{11}$, one can build an analog-digital hybrid computer which requires less power and is 
reasonably small so that the overall computation is basically still done in the analog domain and digital effects will not dominate.
For instance, with $Q$ chosen just as big as $Q=\nu^D/\nu^A$, the analog computer
would evolve only $N/Q=10^7$ points in time, but run $10^4$ times ``in repetition''. The required power
reduces from cluster-grade to desktop-grade $P_A = (N/Q) P_{ND} = 3.3\,\unit{kW}$.
The runtime advantage is of course lost, $T_D/T_A = (Q \nu^A)/\nu^D = 1$.

Naturally, this scenario can also be applied to solve larger problems with a given grid size. For instance,
given an analog computer with the size of $N=10^{11}$ grid points, one can solve a grid of size $Q N$ by
succesively evolving $Q$ parts of the computer with the same power $P_A$ as for a grid of size $N$. Of course, the
overall time to solution and energy will grow with $Q$. In any case, time and energy remain $(3\pm 1)$ orders of
magnitude lower than for a purely digital computer solution.

\conclusions[Summary and outlook]  
\label{sec:summary}

In Section \ref{sec:linear}, we have shown the time and power needs of analog computers
are \emph{orthogonal} to those of digital computers. In Section \ref{sec:ode}, we performed
an actual miniature benchmark of a commercially available 
\emph{Analog Paradigm Model-1} computer versus a mobile Intel\textsuperscript{\textcopyright} processor.
The results are remarkable in several ways: The modern analog computer \emph{Analog Paradigm Model-1},
uses integrated circuit technology which is comparable to the 1970s digital integration level.
Nevertheless it achieves competitive results in computational power and energy consumption compared
to a mature cutting-edge digital processor architecture which has been developed by one of the largest
companies in the world.
We also computed a problem-dependent effective FLOP/sec value for
the analog computer. For the key performance measure for energy-efficient computing,
namely \emph{FLOP-per-Joule}, the analog computer again  obtains remarkable results.

Note that while FLOP/sec is a popular measure in scientific computing, it is always application- and
algorithm-specific. Other measures exist, such as transversed edges per second
(TEPS) or synaptic updates per second (SUPS). \citet{Cockburn2001} propose for instance to measure
the  \emph{efficiency} of a PDE solving method by computing the inverse of the
product of the (spatial-volume integrated) $L^1$-error times the computational cost in
terms of time-to-solution or invested resources.

In Section \ref{sec:pde}, large scale applications were discussed on the example of fluid
dynamics and by comparing high performance computing results with a prospected analog
computer chip architecture. Large scale analog applications can become power-bound and
thus require the adoption of analog-digital hybrid architectures. Nevertheless, with their $\mathcal{O}(1)$
runtime scaling, analog computers excel for time integrating large coupled systems where
algorithmic approaches suffer from communication costs. We predict outstanding advantages in
terms of time-to-solution when it comes to large scale analog computing. Given the advent
of chip-level analog computing, a \emph{gigascale} analog computer (a device with $\sim 10^9$
computing elements) could become a \emph{game changer} in this decade. Of course, major obstacles
have to be addressed to realize such a computer, such as the interconnection toplogy and
realization in an (energy) efficient manner.

Furthermore, there are a number of different approaches in the field of partial differential
equations which might be even better suited to analog computing. For instance, solving
PDEs with artificial intelligence has become a fruitful research field in the last
decade
\citep[see for instance][]{Michoski2020,Schenk2018}, and analog neural networks
might be an interesting candidate to challenge digital approaches.
Number representation on analog computers can be
nontrivial when the dynamical range is large. This is frequently the case with fluid dynamics,
where large density fluctiations are one reason why perturbative solutions fail and numerical
simulations are carried out in the first place. One reason why 
\emph{indirect} alternative approaches such as neural networks could be better suited than
\emph{direct} analog computing networks is that this problem is avoided.
Furthermore, the demand for high accuracy in fluid dynamics can not easily fulfilled by low resolution 
analog computing.  In the end, it is quite possible that
a small-sized analog neural network might outperform a large-sized classical pseudo-linear time evolution
in terms of time-to-solution and energy requirements. Most of these engineering challenges
have not been discussed in this work and are subject to future studies.

\authorcontribution{
	Bernd Ulmann performed the analog simulations. Sven K\"oppel carried out the numerical
	simulations and the estimates. All authors contributed to the article.
} 

\competinginterests{
    There are no competing interests.
} 


\begin{acknowledgements}
  We thank our anonymous referees for helpful comments and corrections.
  We further thank Dr. Chris Giles for many corrections and suggestions which improved
  the text considerably.
\end{acknowledgements}



%
%
%

\bibliographystyle{copernicus}
\bibliography{kht2020}

\begin{thebibliography}{55}
\providecommand{\natexlab}[1]{#1}
\providecommand{\url}[1]{{\tt #1}}
\providecommand{\urlprefix}{URL }
\expandafter\ifx\csname urlstyle\endcsname\relax
  \providecommand{\doi}[1]{https://doi.org/\discretionary{}{}{}#1}\else
  \providecommand{\doi}{https://doi.org/\discretionary{}{}{}\begingroup
  \urlstyle{rm}\Url}\fi

\bibitem[{Amant et~al.(2014)Amant, Yazdanbakhsh, Park, Thwaites, Esmaeilzadeh,
  Hassibi, Ceze, and Burger}]{amant2014}
Amant, R., Yazdanbakhsh, A., Park, J., Thwaites, B., Esmaeilzadeh, H., Hassibi,
  A., Ceze, L., and Burger, D.: General-purpose code acceleration with
  limited-precision analog computation, vol.~42, pp. 505--516,
  \doi{10.1109/ISCA.2014.6853213}, 2014.

\bibitem[{Bishop and Green(1970)}]{bishop1970}
Bishop, K. and Green, D.: Hybrid Computer Impelementation of the Alternating
  Direction Implicit Procedure for the Solution of Two-Dimensional, Parabolic,
  Partial-Differential Equations, AIChE Journal, 16, 139--143, 1970.

\bibitem[{Bournez and Pouly(2018)}]{Bournez2018}
Bournez, O. and Pouly, A.: A Survey on Analog Models of Computation, CoRR,
  abs/1805.05729, \urlprefix\url{http://arxiv.org/abs/1805.05729}, 2018.

\bibitem[{Breems et~al.(2016)Breems, Bolatkale, Brekelmans, Bajoria, Niehof,
  Rutten, Oude-Essink, Fritschij, Singh, and Lassche}]{Breems2016}
Breems, L., Bolatkale, M., Brekelmans, H., Bajoria, S., Niehof, J., Rutten, R.,
  Oude-Essink, B., Fritschij, F., Singh, J., and Lassche, G.: A 2.2 {GHz}
  Continuous-Time Delta Sigma {ADC} With -102 {dBc} {THD} and 25 {MHz}
  Bandwidth, {IEEE} Journal of Solid-State Circuits, 51, 2906--2916,
  \doi{10.1109/jssc.2016.2591826}, 2016.

\bibitem[{Brezis and Browder(1998)}]{Brezis1998}
Brezis, H. and Browder, F.: Partial Differential Equations in the 20th Century,
  Advances in Mathematics, 135, 76--144, \doi{10.1006/aima.1997.1713}, 1998.

\bibitem[{Calude et~al.(1999)Calude, P{\u a}un, T{\u a}t{\u a}r{\^a}m, (a, and
  (b}]{Calude99aglimpse}
Calude, C.~S., P{\u a}un, G., T{\u a}t{\u a}r{\^a}m, M., (a, C. S.~C., and (b,
  G.~P.: A glimpse into natural computing, J. Multi Valued Logic, 7, 2001,
  1999.

\bibitem[{Chu(1979)}]{Yih79}
Chu, C.: Numerical Methods in Fluid Dynamics, vol.~18 of {\em Advances in
  Applied Mechanics\/}, pp. 285--331, Elsevier,
  \doi{10.1016/S0065-2156(08)70269-2}, 1979.

\bibitem[{Cockburn and Shu(2001)}]{Cockburn2001}
Cockburn, B. and Shu, C.-W.: Journal of Scientific Computing, 16, 173--261,
  \doi{10.1023/a:1012873910884}, 2001.

\bibitem[{Cowan et~al.(2005)Cowan, Melville, and Tsividis}]{Cowan2005AVA}
Cowan, G., Melville, R.~C., and Tsividis, Y.~P.: A VLSI analog computer/math
  co-processor for a digital computer, ISSCC. 2005 IEEE International Digest of
  Technical Papers. Solid-State Circuits Conference, 2005., pp. 82--586 Vol. 1,
  2005.

\bibitem[{Cowan(2005)}]{CowanDiss2005}
Cowan, G. E.~R.: A VLSI analog computer/math co-processor for a digital
  computer, Ph.D. thesis, Columbia University, 2005.

\bibitem[{Dahlquist(1979)}]{dahlquist1979generalized}
Dahlquist, G.: Generalized disks of contractivity for explicit and implicit
  Runge-Kutta methods, Tech. rep., CM-P00069451, 1979.

\bibitem[{de~Melo(2010)}]{perf1}
de~Melo, A.~C.: The New Linux {\rq}perf{\rq} Tools, Tech. rep.,
  \urlprefix\url{http://vger.kernel.org/~acme/perf/lk2010-perf-paper.pdf},
  2010.

\bibitem[{Deaton et~al.(1998)Deaton, Garzon, Rose, Franceschetti, and
  Stevens}]{DeatonGarzonRoseFranceschettiStevens1998}
Deaton, R., Garzon, M., Rose, J., Franceschetti, D., and Stevens, S.: DNA
  Computing: A Review, Fundamenta Informaticae, 35, 231--245,
  \doi{10.3233/FI-1998-35123413}, 1998.

\bibitem[{Diot et~al.(2012)Diot, Loub{\`e}re, and Clain}]{Diot2012}
Diot, S., Loub{\`e}re, R., and Clain, S.: The MOOD method in the
  three-dimensional case: Very-High-Order Finite Volume Method for Hyperbolic
  Systems., 2012.

\bibitem[{Dumbser et~al.(2008)Dumbser, Balsara, Toro, and Munz}]{Dumbser2008}
Dumbser, M., Balsara, D.~S., Toro, E.~F., and Munz, C.-D.: A unified framework
  for the construction of one-step finite volume and discontinuous Galerkin
  schemes on unstructured meshes, Journal of Computational Physics, 227,
  8209--8253, \doi{10.1016/j.jcp.2008.05.025}, 2008.

\bibitem[{Fambri et~al.(2018)Fambri, Dumbser, K{\"o}ppel, Rezzolla, and
  Zanotti}]{Fambri2018}
Fambri, F., Dumbser, M., K{\"o}ppel, S., Rezzolla, L., and Zanotti, O.: {ADER}
  discontinuous Galerkin schemes for general-relativistic ideal
  magnetohydrodynamics, Monthly Notices of the Royal Astronomical Society,
  \doi{10.1093/mnras/sty734}, 2018.

\bibitem[{Feilmeier(1974)}]{feilmeier1974}
Feilmeier, M.: Hybridrechnen, Springer, \doi{10.1007/978-3-0348-5490-0}, 1974.

\bibitem[{Georgescu et~al.(2014)Georgescu, Ashhab, and Nori}]{Georgescu_2014}
Georgescu, I.~M., Ashhab, S., and Nori, F.: Quantum simulation, Reviews of
  Modern Physics, 86, 153--185, \doi{10.1103/revmodphys.86.153}, 2014.

\bibitem[{Gruber et~al.(2020)Gruber, Eitzinger, Hager, and Wellein}]{LikwidDoi}
Gruber, T., Eitzinger, J., Hager, G., and Wellein, G.: LIKWID 5: Lightweight
  Performance Tools, \doi{10.5281/zenodo.4275676}, 2020.

\bibitem[{Gustafson(1988)}]{Gustafson1988}
Gustafson, J.~L.: Reevaluating Amdahl{\textquotesingle}s law, Communications of
  the {ACM}, 31, 532--533, \doi{10.1145/42411.42415}, 1988.

\bibitem[{Hager et~al.(2010)Hager, Wellein, and Treibig}]{Likwid}
Hager, G., Wellein, G., and Treibig, J.: LIKWID: A Lightweight
  Performance-Oriented Tool Suite for x86 Multicore Environments, in: 2012 41st
  International Conference on Parallel Processing Workshops, pp. 207--216, IEEE
  Computer Society, Los Alamitos, CA, USA, \doi{10.1109/ICPPW.2010.38}, 2010.

\bibitem[{Harten(1997)}]{harten1997high}
Harten, A.: High resolution schemes for hyperbolic conservation laws, Journal
  of computational physics, 135, 260--278, 1997.

\bibitem[{Hirsch(1990)}]{hirsch1990numerical}
Hirsch, C.: Numerical computation of internal and external flows. Vol.
  2-Computational Methods for Inviscid and Viscous Flows, Chichester, 1990.

\bibitem[{Huang et~al.(2017)Huang, Guo, Seok, Tsividis, Mandli, and
  Sethumadhavan}]{Huang2017}
Huang, Y., Guo, N., Seok, M., Tsividis, Y., Mandli, K., and Sethumadhavan, S.:
  Hybrid analog-digital solution of nonlinear partial differential equations,
  in: Proceedings of the 50th Annual {IEEE}/{ACM} International Symposium on
  Microarchitecture, {ACM}, \doi{10.1145/3123939.3124550}, 2017.

\bibitem[{Karplus and Russell(1971)}]{karplus1971}
Karplus, W. and Russell, R.: Increasing Digital Computer Efficiency with the
  Aid of Error-Correcting Analog Subroutines, IEEE Transactions on Computers,
  C-20, 1971.

\bibitem[{Kendon et~al.(2010)Kendon, Nemoto, and Munro}]{Kenden2010}
Kendon, V.~M., Nemoto, K., and Munro, W.~J.: Quantum analogue computing,
  Philosophical Transactions of the Royal Society A: Mathematical, Physical and
  Engineering Sciences, 368, 3609--3620, \doi{10.1098/rsta.2010.0017}, 2010.

\bibitem[{K{\"o}ppel(2018)}]{Koeppel2018}
K{\"o}ppel, S.: Towards an exascale code for {GRMHD} on dynamical spacetimes,
  Journal of Physics: Conference Series, 1031, 012\,017,
  \doi{10.1088/1742-6596/1031/1/012017}, 2018.

\bibitem[{MacLennan(2004)}]{MacLennan2004}
MacLennan, B.~J.: Natural computation and non-Turing models of computation,
  Theoretical Computer Science, 317, 115--145, \doi{10.1016/j.tcs.2003.12.008},
  super-Recursive Algorithms and Hypercomputation, 2004.

\bibitem[{MacLennan(2012)}]{MacLennan2012}
MacLennan, B.~J.: Analog Computation, in: Computational Complexity, pp.
  161--184, Springer New York, \doi{10.1007/978-1-4614-1800-9_12}, 2012.

\bibitem[{MacLennan(2019)}]{MacLennan2019}
MacLennan, B.~J.: Unconventional Computing, University of Tennessee,
  \urlprefix\url{http://web.eecs.utk.edu/~bmaclenn/Classes/494-594-UC/handouts/UC.pdf},
  2019.

\bibitem[{Michoski et~al.(2020)Michoski, Milosavljevi{\'c}, Oliver, and
  Hatch}]{Michoski2020}
Michoski, C., Milosavljevi{\'c}, M., Oliver, T., and Hatch, D.~R.: Solving
  differential equations using deep neural networks, Neurocomputing, 399,
  193--212, \doi{10.1016/j.neucom.2020.02.015}, 2020.

\bibitem[{Nomura and Deiters(1968)}]{nomura1968}
Nomura, T. and Deiters, R.: Improving the analog simulation of partial
  differential equations by hybrid computation, Simulation, 1968.

\bibitem[{Reihing(1959)}]{reihing1959}
Reihing, J.: A time-sharing analog computer, in: Proceedings of the western
  joint computer conference, 1959.

\bibitem[{Rodgers(1985)}]{AmdahlsLaw1985}
Rodgers, D.~P.: Improvements in Multiprocessor System Design, SIGARCH Comput.
  Archit. News, 13, 225--231, \doi{10.1145/327070.327215}, 1985.

\bibitem[{R{\"o}hl et~al.(2017)R{\"o}hl, Eitzinger, Hager, and
  Wellein}]{Likwid2}
R{\"o}hl, T., Eitzinger, J., Hager, G., and Wellein, G.: {LIKWID} Monitoring
  Stack: {A} flexible framework enabling job specific performance monitoring
  for the masses, CoRR, abs/1708.01476,
  \urlprefix\url{http://arxiv.org/abs/1708.01476}, 2017.

\bibitem[{Schenck and Fox(2018)}]{Schenk2018}
Schenck, C. and Fox, D.: SPNets: Differentiable Fluid Dynamics for Deep Neural
  Networks, CoRR, abs/1806.06094,
  \urlprefix\url{http://arxiv.org/abs/1806.06094}, 2018.

\bibitem[{Schuman et~al.(2019)Schuman, Potok, Patton, Birdwell, Dean, Rose, and
  Plank}]{Schuman.5192017}
Schuman, C.~D., Potok, T.~E., Patton, R.~M., Birdwell, J.~D., Dean, M.~E.,
  Rose, G.~S., and Plank, J.~S.: A Survey of Neuromorphic Computing and Neural
  Networks in Hardware, \urlprefix\url{http://arxiv.org/pdf/1705.06963v1},
  2019.

\bibitem[{Shu(2016)}]{SHU2016598}
Shu, C.-W.: High order WENO and DG methods for time-dependent
  convection-dominated PDEs: A brief survey of several recent developments,
  Journal of Computational Physics, 316, 598--613,
  \doi{10.1016/j.jcp.2016.04.030}, 2016.

\bibitem[{Siegelmann(1995)}]{Siegelmann1995}
Siegelmann, H.~T.: Computation Beyond the Turing Limit, Science, 268, 545--548,
  \doi{10.1126/science.268.5210.545}, 1995.

\bibitem[{Sod(1985)}]{Sod1985}
Sod, G.: Numerical Methods in Fluid Dynamics: Initial and Initial
  Boundary-Value Problems, Cambridge University Press, 1985.

\bibitem[{Subramaniam et~al.(2013)Subramaniam, Saunders, Scogland, and
  Feng.}]{green500_1}
Subramaniam, B., Saunders, W., Scogland, T., and Feng., W.-c.: Trends in
  Energy-Efficient Computing: A Perspective from the Green500, in: Proceedings
  of the International Green Computing Conference, 2013.

\bibitem[{Subramaniam et~al.(2020)}]{green500_2}
Subramaniam, B. et~al.: Green 500 List, 2020, see {http://www.green500.org},
  2020.

\bibitem[{Titarev and Toro(2005)}]{TITAREV2005715}
Titarev, V. and Toro, E.: ADER schemes for three-dimensional non-linear
  hyperbolic systems, Journal of Computational Physics, 204, 715--736,
  \doi{10.1016/j.jcp.2004.10.028}, 2005.

\bibitem[{Titarev and Toro(2002)}]{Titarev2002}
Titarev, V.~A. and Toro, E.~F.: Journal of Scientific Computing, 17, 609--618,
  \doi{10.1023/a:1015126814947}, 2002.

\bibitem[{Toro(1998)}]{Toro1998}
Toro, E.~F.: Primitive, Conservative and Adaptive Schemes for Hyperbolic
  Conservation Laws, in: Numerical Methods for Wave Propagation, pp. 323--385,
  Springer Netherlands, \doi{10.1007/978-94-015-9137-9_14}, 1998.

\bibitem[{Ulmann(2019)}]{Model1Handbook}
Ulmann, B.: Model-1 Analog Computer Handbook/User Manual,
  \urlprefix\url{http://analogparadigm.com/downloads/handbook.pdf}, 2019.

\bibitem[{Ulmann(2020)}]{ap2}
Ulmann, B.: Analog and Hybrid Computer Programming, De Gruyter, 2020.

\bibitem[{Vichnevetsky(1968)}]{vichnevetsky1968}
Vichnevetsky, R.: A new stable computing method for the serial hybrid computer
  integration of partial differential equations, in: Spring Joint Computer
  Conference, 1968.

\bibitem[{Vichnevetsky(1971)}]{vichnevetsky1971}
Vichnevetsky, R.: Hybrid methods for partial differential equations,
  Simulation, 1971.

\bibitem[{Volynskii and Bukham(1965)}]{volynskii1965}
Volynskii and Bukham: Analogues for the Solution of Boundary-Value Problems,
  Pergamon Press, library of Congress Catalog Card No 64-25643, 1965.

\bibitem[{Wang et~al.(2018)Wang, Zhu, Chan, and Martins}]{Wang2018}
Wang, W., Zhu, Y., Chan, C.-H., and Martins, R.~P.: A 5.35-{mW} 10-{MHz}
  Single-Opamp Third-Order {CT} Delta Sigma Modulator With {CTC} Amplifier and
  Adaptive Latch {DAC} Driver in 65-nm {CMOS}, {IEEE} Journal of Solid-State
  Circuits, 53, 2783--2794, \doi{10.1109/jssc.2018.2852326}, 2018.

\bibitem[{Wang et~al.(2019)Wang, Yu, Berto, Cai, and Bao}]{Yi19}
Wang, Y., Yu, B., Berto, F., Cai, W., and Bao, K.: Modern numerical methods and
  theirapplications in mechanical engineering, Advances in Mechanical
  Engineering, 11, \doi{10.1177/1687814019887255}, 2019.

\bibitem[{Wilhelm et~al.(2017)Wilhelm, Steinwandt, Langenberg, Liebermann,
  Messinger, and Schuhmacher}]{BSIquantenComputer}
Wilhelm, F., Steinwandt, R., Langenberg, B., Liebermann, P., Messinger, A., and
  Schuhmacher, P.: Status of quantum computer development, 2017.

\bibitem[{Zhou et~al.(2020)Zhou, Stoudenmire, and Waintal}]{Zhou_2020}
Zhou, Y., Stoudenmire, E.~M., and Waintal, X.: What Limits the Simulation of
  Quantum Computers?, Physical Review X, 10, \doi{10.1103/physrevx.10.041038},
  2020.

\bibitem[{Ziegler(2020)}]{Ziegler2020}
Ziegler, M.: Novel hardware and concepts for unconventional computing, Sci Rep,
  10, \doi{10.1038/s41598-020-68834-1}, 2020.

\end{thebibliography}

\end{document}